\documentstyle[12pt,epsfig]{article}
\begin{document}

\vspace*{2.cm}
\begin{center}
{\bf Lepton Asymmetry and Neutrino Oscillations Interplay}

\vspace*{0.2cm}
Daniela Kirilova$^{\dagger}$\\[0.3cm]
{\it $^\dagger$Institute of Astronomy and NAO, Bulgarian Academy of
Sciences, Sofia}
\end{center}
\vspace*{0.2cm}
\begin{abstract}
 We discuss the interplay between lepton asymmetry L and $\nu$ oscillations
 in the early Universe. Neutrino oscillations may suppress or enhance previously
 existing L. On the other hand L is capable to suppress or enhance neutrino
 oscillations. The mechanism of L enhancement in MSW resonant $\nu$ oscillations
 in the early Universe is numerically analyzed.
  L  cosmological effects through $\nu$ oscillations are discussed.
  We discuss how L  may change the cosmological BBN constraints on
neutrino and show that BBN model with $\nu_e \leftrightarrow \nu_s$
oscillations is extremely sensitive to  L  - it allows to obtain the
most stringent constraints on L value.
 We discuss also the cosmological role of active-sterile $\nu$ mixing and L
 in connection with the indications about additional relativistic density in
 the early Universe, pointed out by BBN, CMB and LSS data and the analysis of global $\nu$ data.

\ \\
keywords: lepton asymmetry, neutrino oscillations, BBN, excess
radiation

\end{abstract}

\section*{Introduction}

Lepton asymmetry (L) of the Universe is not measured yet and may be
orders of magnitude bigger than the baryon asymmetry $\beta \sim
6.10^{-10}$, which was measured with great precision from Cosmic
Microwave Background (CMB) and Big Bang Nucleosynthesis (BBN) data.
L is usually defined as $L=(N_l-N_{\bar{l}})/N_{\gamma}$, where
$N_l$ is the number density of leptons, $N_{\bar{l}}$ of
antileptons, while $N_{\gamma}$ is the number density of photons.
Considerable L might be contained only in $\nu$ sector. Thus, the
detection of the Cosmic Neutrino Background would provide L direct
measurement. Till then L is measured indirectly by its influence on
observable relics  of the Universe. The abundances of the
primordially produced light elements during BBN provide such a
sensitive test of L.

Main cosmological effects of L on BBN  are its dynamical and kinetic
effect. {\it The dynamical effect of L} consists in the increase of
the radiation energy density
 of the Universe due to non-zero L:  $\delta \rho_r = [1+7/8(4/11)^{4/3}\delta
 N_{eff}]\rho_{\gamma}$, where
 $\Delta N_{eff}=15/7[(\xi/\pi)^4+2(\xi/\pi)^2]$, where $\xi=\mu/T$ is the
 $\nu$ degeneracy parameter. This effect leads to faster Universe
expansion $H=(8/3\pi G \rho)^{1/2}$, which leads to earlier freezing
of nucleons and hence influences BBN produced elements. {\it The
kinetic effect} is noticeable for big enough $|L|>0.01$ in the
$\nu_e$ sector and is due to different number densities of $\nu_e$
and $\bar{\nu_e}$, which participates in neutron-proton transfers:
$\nu_e + n \leftrightarrow p + e^-$, $\bar{\nu_e} + p
\leftrightarrow n + e^+$ in the pre-BBN epoch and, correspondingly,
BBN yields (see ref.~\cite{Simha&Steigman} and refs there in).

Besides these well known effects L may influence BBN through its
interplay with neutrino oscillations. Unlike previous effects, this
L effect on BBN, called further {\it indirect kinetic effect},  may
be noticeable for much smaller L values, $L<<0.01$~\cite{NPB98}.
Hence BBN with oscillations provides the possibility to measure
and/or constrain tiny L with values close to $\beta$.

In the next section the interplay between L  and $\nu$ oscillations
is described. In the third section the indirect kinetic effect of L
on BBN is studied and the possibility to constrain L via BBN with
$\nu$ oscillations is discussed.  L  role as a solution of the
problem of the excess radiation density in the Universe is provided
in the fourth section.

\section{Interplay between  L  and neutrino oscillations in the early Universe}

  Small $L<<0.01$  has
negligible dynamic and direct kinetic effect. Nevertheless, due to
its interplay with $\nu$ oscillations such L is capable of
         changing $\nu$  number densities,
         $\nu$  distribution and spectrum
         distortion and
         changing $\nu$ oscillations pattern (suppressing or enhancing
         oscillations), by which it influences nucleons
kinetics and finally BBN production of light elements~\cite{NPB98}.
This effect was proven to persist
 down to $L \sim 10^{-8}$~\cite{NPB98,PPNP11,JCAP12}.

We studied numerically the interplay between tiny L,
$10^{-10}<L<10^{-4}$, and electron-sterile $\nu_e \leftrightarrow
\nu_s$ oscillations,
 effective after active
$\nu$ decoupling $\delta m^2 \sin^4 2\theta \le 10^{-7}$ eV$^2$.

{\it It is known that active-sterile oscillations}  may  {\it change
   neutrino-antineutrino asymmetry
 of the medium}, suppress or enhance it
 ~\cite{NPB98,PPNP11,JCAP12,Dolgov81,BarbieriDolgov90,KC,FV95,FV97,shi,NPB00,DolgovVillante00,FV96,D04,Bari}.
 ~\footnote{There are other cosmological effects of active-sterile
neutrino oscillations, like excitation of additional light particles
into equilibrium~\cite{Dolgov81,BarbieriDolgov90} and  distortion
$\nu$ energy distribution~\cite{DK88,KC}.} On the other hand L
influences $\nu$ propagation. Qualitatively this influence may be
described as follows: The average potentials $V_f$ for $\nu$ depend
on the particle asymmetries of different constituents of the medium
and they differ for different neutrino types due to different
interactions with the particles of the
plasma~\cite{Notzold&Raffelt88}:

\begin{equation}
V_f=\sqrt{2} G_F QN_{\gamma}/M_W^2 \pm L N_{\gamma}
\end{equation}

 \noindent where
$f=e, \mu, \tau$, "minus"corresponds to $\nu$ and "plus" to
$\bar{\nu}$, $Q\sim-ET$,       $L\sim-L_{\alpha}$,
                    $L^{\alpha}$ is given through the
fermion asymmetries of the plasma (in the discussed case $L \sim
2L_{\nu_e} + L_{\nu_{\mu}} + L_{\nu_{\tau}}$).

  In the adiabatic case the effect of the medium can be hidden in the oscillation parameters
$\delta m^2$ and $\vartheta$ by introducing matter oscillation
parameters:
\begin{equation}
\sin^2\vartheta_m=\sin^2\vartheta/[\sin^2\vartheta+2G_F((Q/M_W^2 \mp
L)N_{\gamma}/\delta m^2-\cos2\vartheta)^2].
\end{equation}
 L, as a characteristic of the  medium, may suppress oscillations by
decreasing their amplitude, or enhance oscillation transfer in case
resonant condition between the parameters of the medium and the
oscillation parameters holds:
\begin{equation}
2G_F(Q/M_W^2 \mp L)N_{\gamma}=\cos2\vartheta \times \delta m^2
\end{equation}
 In the early Universe at high temperature  $Q/M_W^2>L$  resonant oscillations
 both for $\nu$ and $\bar{\nu}$ are
possible if $\delta m^2 <0$.  With the cooling of the Universe when
L begins to dominate, $Q/M_W^2<L$, resonant transfer for
antineutrinos in case $\delta m^2 <0$, or for neutrinos if $\delta
m^2>0$ is possible.~\footnote{The resonant condition (3) for $Q=0$
was first studied and  is known as Mikheev-Smirnov-Wolfenstein
effect~\cite{MSW}.} Thus neutrino propagation and resonance in the
neutrino sector differs from that of antineutrino for non-zero L.
Due to  L influence of the $\nu$ propagation L may change $n_{\nu}$,
its spectrum distribution and oscillation pattern.

This simplified description of the medium influence  is applicable
in the equilibrium situation, when working in terms of average $\nu$
momentum and particle densities is reasonable. In the nonequilibrium
situation, when spectrum distribution of $\nu$ is considerable, as
is the case of late electron-sterile oscillations discussed here, it
strongly effects both $\nu$ propagation and L evolution. Hence, for
the correct description of the neutrino - asymmetry interplay it is
essential to provide an accurate account of the neutrino spectrum
distortion due to L and oscillations.
 ~\footnote{For example, in the nonequilibrium situation, when spectrum
distribution of $\nu$ was properly described, simultaneous resonance
transfer was found possible also in the $\delta m^2 <0$ case due to
the "resonant wave" passing through the neutrino distribution
~\cite{NPB98}.}

\subsection{Exact description of the propagation of neutrinos and L evolution}

The equations governing  $\nu$ evolution, in terms of neutrino
density matrix in momentum space, are given below. They account
simultaneously for neutrino-L interplay, Universe expansion, $\nu$
oscillations and $\nu$ forward scattering and describe precisely
$\nu$ energy distribution~\cite{KC}. Our numerical analysis of L and
$\nu$ propagation was based on these equations:

$$
\partial \rho(t) / \partial t =
H p_\nu~ \left(\partial \rho(t) / \partial p_\nu\right) +$$
$$ + i
\left[ {\cal H}_o, \rho(t) \right] +i \sqrt{2} G_F \left({\cal L} -
Q/M_W^2 \right)N_\gamma \left[ \alpha, \rho(t) \right] + {\rm
O}\left(G_F^2 \right)
$$
$$
\partial\bar\rho(t) / \partial t=
H p_\nu~ \left(\partial \bar\rho(t) / \partial p_\nu\right) +
$$
 $$
+ i \left[ {\cal H}_o,\bar\rho(t) \right] +i \sqrt{2} G_F
\left(-{\cal L} - Q/M_W^2 \right)N_\gamma \left[ \alpha, \bar\rho(t)
\right] + {\rm O}\left(G_F^2 \right).
$$
$$
L_{\nu_e} \sim \int {\rm d}^3p(\rho_{LL}-\bar{\rho}_{LL})/N_\gamma
$$
\ \\

\noindent where $\alpha_{ij}=U^*_{ie} U_{je}$,
 $\nu_i=U_{il}\nu_l (l=e,s)$. ${\cal H}_o$ is the free $\nu$ Hamiltonian.
 $Q$ arises as an $W/Z$ propagator effect,$Q \sim E_\nu~T$.
 {${\cal L} \sim 2L_{\nu_e}+L_{\nu_\mu}+L_{\nu_\tau}$,
$L_{\mu,\tau} \sim (N_{\mu,\tau}-N_{\bar{\mu},\bar{\tau}})/
N_\gamma$. At decoupling of  $\nu_e$  $\nu_s$ was assumed
empty.~\footnote{ The case of non-zero population of $\nu_s$  was
considered in refs.~\cite{D04,D07,DP06}.}

Due to $L$ the equations  are coupled integro-differential. $L$
leads to different evolution of $\nu$ from $\bar{\nu}$ due to the
different sign of L in the equations. Numerical analysis of the
evolution of  $\nu$ ensembles, evolution of $L$, and also the
evolution of nucleons for the entire range of oscillation parameters
and for the temperature range [$2$ MeV, $0.3$ MeV]  and for
$10^{-10}<L<0.01$ was provided in non-resonant $\nu$ oscillations
case. In case of resonant oscillations
 L initial value was taken to be $L_i \sim \beta$.
We have described precisely  $\nu$ momenta distribution:  5000 bins
were used in the non-resonant oscillations case, and up to 10 000 in
the resonant case.

  In case of {\it nonresonant oscillations and relic L}
  the following relations describe with good accuracy the exact behavior of
$L$ and  $L$-oscillations interplay:
    $L \ge 10^{-7}$  enhances oscillations,  $L>0.1 (\delta
     m^2/{\rm eV}^2)^{2/3}$
    suppresses oscillations, and  asymmetries
    $L>(\delta m^2/{\rm eV}^2)^{2/3}$
    inhibit oscillations.
    For illustration of the exact dependence see the figures in ref. ~\cite{JCAP12}.

 In case of {\it resonant} $\nu_e \leftrightarrow \nu_s$
{\it oscillations} the evolution of $L$ has a rapid oscillatory
behavior.  The region of parameter space for which a generation of L
is possible was found $|\delta m^2|\sin^4 2\theta \le 10^{-9.5}$
eV$^2$. ~\footnote{The instability region is slightly more stringent
than the existing in literature for other oscillation models
ref.~\cite{dolgL}.} A maximum possible growth of $L$ by 5 orders of
magnitude was determined.

 L role in BBN with $\nu$ oscillations was numerically studied as well.
 The change in BBN constraints on oscillation parameters due to L and BBN constraints on L
in case of $\nu_e \leftrightarrow \nu_s$ are presented in the next
section.

\section{BBN with active-sterile neutrino oscillations and lepton asymmetry}

Big Bang Nucleosynthesis  is theoretically well established due to
the precise data on nuclear processes rates, known from laboratory
experiments at the energies relevant to BBN epoch, 
and the precise data on the light elements abundances D, He and Li.
Besides, the baryon-to-photon ratio, which is the only parameter in
the standard BBN, has been independently measured with good accuracy
by CMB precision data. Therefore, BBN is used as the most early and
precision probe for physical conditions in early Universe and,
hence, presents the best astrophysical and cosmological probe for
new physics and microphysics, relevant  at BBN energies.

        Being the best speedometer at radiation dominated stage
                 BBN was used to probe $\nu$ properties, the number of light species,
                  $\nu$ oscillations, distortion in $\nu$ distribution, etc.
                  Due to its sensitivity to the expansion rate and to  the
                  nucleons kinetics in the pre-BBN epoch it has been
                  shown that BBN presents also
         the most exact leptometer (see ref.~\cite{JCAP12} and
         references there in).
Thus BBN studies allow to put stringent limits on oscillation
parameters~\cite{Dolgov81,DK88,BarbieriDolgov90,KC,PRD,NPB00,DolgovVillante00,D04,D07,DP06,Panayotova11}
and on L in the presence of electron-sterile
oscillations~\cite{NPB98,PPNP11,JCAP12}.

\subsection{BBN constraints on L}

Eventual $\nu$  degeneracies of flavor neutrinos will equilibrate
before BBN due to flavor $\nu$ oscillations, having in mind recently
measured value  $\theta_{13}$. Then  BBN constraints on L read
~\cite{DolgovPetcov,ABB02,Iocco09,Serpico&Raffelt,PastorPinto&Raffelt,mangano}:
\begin{equation}
|L|<0.1.
\end{equation}

 These constraints allow the possibility of  L orders of
 magnitude larger than the measured $\beta$. However, for such small $L$ the dynamical cosmological
 effect of $L$ isnegligible.
Hence, that small L in the flavor $\nu$ sector cannot mimic the
extra relativistic degrees of freedom during BBN, which seem to be
required by  recent analysis of  cosmological and neutrino
oscillations data (to be discussed in more detail in the fourth
section).

\subsection{L and BBN with late electron sterile $\nu$ oscillations}

 In case of $L\sim \beta$ stringent BBN constraints on active-sterile oscillation parameters exist.
{\it The presence of}  $L>\beta$ {\it may change considerably BBN
constraints on oscillations parameters} due to L indirect kinetic
effect.

In BBN with electron-sterile $\nu$ oscillations $Y_p$ decreases at
small mixing parameters values due to L growth caused by resonant
$\nu$ transfer, thus at these mixing angles {\it oscillations
generated L relaxes the BBN constraints on
oscillations}~\cite{KC,NPB00,JCAP12}.

 In case of relic $L>\beta$
the presence of such L relaxes the BBN bounds on neutrino
oscillation parameters in case of electron-sterile non-resonant
neutrino oscillations at maximal mixing and strengthens them at
small mixing angles. In this case depending on the
asymmetry-oscillations interplay the asymmetry may enhance, suppress
or stop $\nu$ oscillations, reflecting correspondingly to
strengthening, relaxation or elimination of the BBN constraints. In
the last case the approximate BBN constraint reads:
\begin{equation}
\delta m^2 (eV^2) < L^{3/2}.
\end{equation}

Vice versa, this constraint can be considered also as a cosmological
constraint on L from BBN with electron-sterile neutrino
oscillations. Thus, L generated in the electron-sterile sector may
(partially) suppress the oscillations in other sectors, hindering
equalization of chemical potentials, thus relaxing the stringent BBN
bound. The presence of L (no matter how generated), capable to
suppress oscillations, may lead to only partial population of the
sterile neutrino. Vise versa L values capable to enhance
oscillations will bring faster the sterile state into equilibrium.

The model of BBN with $\nu_e \leftrightarrow \nu_s$, effective after
electron $\nu$ decoupling presents the possibility to feel extremely
small asymmetries $L>10^{-8}$ due to the indirect kinetic effect of
L (see ref.~\cite{JCAP12}). This sensitivity allows to derive
stronger constraints on $L$. Namely, equation (5) can be used to put
the following BBN constraint on L: Having the indications for
active-sterile oscillations with $\delta m^2 \sim
10^{-5}$~\cite{HolandaSmirnov} and replacing this value in  eq.(5),
a much stronger  upper limit on L  (than the one presented in
eq.(4)) follows $L<10^{-3.3}$.

\section{L and excess radiation density in the Universe}

In recent years an increasing number of {\it cosmological
indications} suggesting excess relativistic density, corresponding
to different epochs,
appeared~\cite{IT10,Aver10,WMAP7,Keisler11,Dunkley11,hou11,riemer}.
~\footnote{See, however, the recent results~\cite{bennett}, which
are in agreement with the canonical value of $N_{eff}$ and also the
discussion in ref.\cite{abazajan}.} Besides, {\it $\nu$ oscillations
data} require 1 or 2 additional sub-eV sterile neutrino,
participating into oscillations with flavor neutrinos with higher
mass differences values, than the ones required by solar and
atmospheric $\nu$ oscillations experiments
~\cite{Kopp11,GiuntiLaveder10,AkhmedovSchwetz10,RazzaqueSmirnov11,mention11}.
It is interesting if cosmology allows 2 light additional sterile
neutrinos and if they can explain the excess relativistic density.

 Additional light sterile neutrinos with the  mixing and mass
differences estimated by $\nu$ oscillations data with mass
differences in the eV range will be brought into equilibrium in the
early Universe. BBN favors the presence of one such $\nu_s$ but He
and D data excludes 2 fully thermalized $\nu_s$. Besides,  neutrinos
in sub-eV range produce too much hot dark
matter~\cite{Hamann10,dodelson06}. Thus, two additional $\nu_s$ are
in tension both with BBN and with LSS
requirements~\cite{Giusarma11,NolletHolder11,hou11}.

 L, namely its dynamical and direct kinetic effects, has been considered
 as an explanation of the excess radiation. It was shown that
excess radiation cannot be explained by degenerate
BBN~\cite{mangano}. However, the presence of  L  may be the solution
in case its value is enough to
 suppress active-sterile oscillations  so that $\nu_s$ are not fully
 thermalized~\cite{BarbieriDolgov90,FV95,NPB98,hannestad12}.
  Our estimation of the value of L necessary to suppress oscillations
 and achieve the suppression of  $\nu_s$ production  is $L \ge 0.08$. This is higher
 than the values discussed by refs.\cite{hannestad12,mirizzi}, that found $|L|>10^{-2}$.
 The difference might be due to different approximations used or to
 the fact that previous studies do not account precisely for the $\nu$ energy
 distribution.

Thus, in modified BBN with $\nu$ oscillations and high enough L the
models with additional light sterile neutrinos may be allowed. To
obtain the exact L value a precise numerical
 analysis, solving the exact kinetic equations, including all $\nu$ species and
 accounting for all L effects, discussed above, should be
 provided.

 Hence, the excess relativistic density might point to additional sterile
 neutrinos and the presence of L.
  However, there exist other possibilities as well, namely:
      $\nu$ active-sterile late oscillations leading to the overproduction of He-4
      and thus imitating extra radiation,  MeV decaying particles during BBN~\cite{decay}, or
       other modifications of the
      standard cosmological model.
       Future experimental and observational data will choose among different
       possibilities. In particular, it is expected that
       Planck data will be able to check with higher sensitivity the status of extra radiation.

\section{Conclusions}

There exists interesting interplay between small  L, relic or
produced by active-sterile $\nu$ oscillations, and late
active-sterile $\nu$ oscillations. A detail numerical analysis of
this interplay between lepton asymmetry
       $L << 0.01$,
       and $\nu$ oscillations was carried out. The evolution of $\nu$ and L  was studied
       using exact kinetic equations for $\nu$ density matrix in momentum space and
       describing the $\nu$ energy distribution with very high accuracy -
       up to 10 000 bins.

 A considerable enhancement of  L  - by 5
orders of magnitude - was found possible in non-equilibrium
electron-sterile resonant neutrino  oscillations, effective after
the decoupling of active neutrino. The region in the oscillation
parameter space of considerable  L growth was determined.

 The parameter ranges for which relic L  is able to enhance, suppress or inhibit
 non-resonant electron-sterile oscillations was determined.

 Cosmological influence of  small L, which do not have direct effect on nucleons kinetics
 during BBN, was discussed. Such small asymmetries are invisible by CMB at present, but may be felt by
 BBN.  It was shown that L as
small as $10^{-8}$ may be felt by BBN via oscillations. Also  BBN
constraints on $\nu$ oscillations parameters depend nontrivially on
L.

 L generated by oscillations at small mixing angles suppresses
$\nu$ oscillations and reduces the overproduction of $Y_p$ and
relaxes BBN constraints.

Relic L present during BBN, depending on its value, can strengthen,
relax or wave out BBN constraints on oscillations.  The value of L,
capable to hinder $\nu$ oscillations is determined for different
sets of oscillations parameters of the model, a good approximation
to the exact value is $L>(\delta m^2/(eV^2))^{2/3}$. On the other
hand this can be considered as the most  stringent cosmological
constraint  on L based on BBN  with electron-sterile neutrino
oscillations.

Due to its interplay with neutrino oscillations L may play important
role for resolving the problem of additional radiation density
suggested by cosmological observational data and by $\nu$
oscillations data.
       In principal  BBN with L may allow 3+1 and 3+2 oscillations models, in case  L
       value is high enough to
        suppress active-sterile oscillations, thus providing incomplete thermalization of
        $\nu_s$ and relaxing BBN constraints on additional light sterile species.

{\bf Acknowledgements.} I am glad to thank Theoretical and
Computational Physics and Astrophysics Foundation, Sofia University,
for the financial support of my travel. I acknowledge the financial
support by the conference for my participation.  I thank the unknown
referee for the suggestions and criticism that helped to improve the
paper.

\end{document}